\documentclass{aastex}
\usepackage{amsmath}
\usepackage{amsfonts}
\usepackage{amssymb}
\usepackage{graphicx}
\shorttitle{Planetesimal erosion during the JEB}
\shortauthors{Turrini, Coradini \& Magni} 
\begin{document}
\title{Jovian Early Bombardment: planetesimal erosion in the inner asteroid belt}
\author{D. Turrini, A. Coradini, G. Magni}
\affil{Istituto di Astrofisica e Planetologia Spaziali, INAF-IAPS, Via Fosso del Cavaliere 100, 00133, Rome, Italy}
\email{diego.turrini@ifsi-roma.inaf.it}
%
\begin{abstract}
The asteroid belt is an open window on the history of the Solar System, as it preserves records of both its formation process and its secular evolution. The progenitors of the present-day asteroids formed in the Solar Nebula almost contemporary to the giant planets. The actual process producing the first generation of asteroids is uncertain, strongly depending on the physical characteristics of the Solar Nebula, and the different scenarios produce very diverse initial size-frequency distributions. In this work we investigate the implications of the formation of Jupiter, plausibly the first giant planet to form, on the evolution of the primordial asteroid belt. The formation of Jupiter triggered a short but intense period of primordial bombardment, previously unaccounted for, which caused an early phase of enhanced collisional evolution in the asteroid belt. Our results indicate that this Jovian Early Bombardment caused the erosion or the disruption of bodies smaller than a threshold size, which strongly depends on the size-frequency distribution of the primordial planetesimals. If the asteroid belt was dominated by planetesimals less than $100$ km in diameter, the primordial bombardment would have caused the erosion of bodies smaller than $200$ km in diameter. If the asteroid belt was instead dominated by larger planetesimals, the bombardment would have resulted in the destruction of bodies as big as $500$ km. 
\end{abstract}
\section{Introduction}
The asteroid belt is a window open on the past of the Solar System, as it contains records of both its formation process and its secular evolution. However, disentangling the sequence of events that characterized its history is a difficult task, since the more ancient features have been erased or masked by the more recent ones (see e.g. \citet{cea11} and \citet{oas11} for an in-depth discussion). We know from meteoritic constrains (see e.g. \citet{sco07} for a review) and theoretical models (see e.g. \citet{wei11}) that the progenitors of the present-day asteroids formed  and, in some cases, differentiated in the Solar Nebula on a $1$ Ma time-scale, soon followed by Jupiter and the other giant planets (see e.g. \citet{sco06}). However, the actual scenario for the formation of the planetesimals, its efficiency and the resulting initial size-frequency distribution are still poorly constrained. The proposed formation scenarios differ in the assumptions on the formation environment, i.e. a quiescent or a turbulent Solar Nebula, and produce very diverse size-frequency distributions (SFDs in the following) of the primordial planetesimals (see e.g. \citet{cea11} and \citet{oas11} and references therein). The number of planetesimals populating the asteroid belt at this early time was likely $2-3$ orders of magnitude higher than the present population of asteroids \citep{wei77}, so the collisional evolution of the primordial asteroid belt was characterized by a more rapid pace than the present one \citep{bea05a,bea05b}. According to recent theoretical results \citep{mea09,wei11}, in a few Ma the size distribution of the planetesimals in the asteroid belt spanned over $7$ orders of magnitude, covering the range between $10^{-2}$ km and $10^{4}$ km independently on the formation mechanism.\\
The present view of the evolution of the asteroid belt assumes that it underwent a major phase of depletion and enhanced collisional evolution after the dispersal of the nebular gas due to the combined perturbations of planetary embryos in the inner Solar System and of the giant planets in the outer Solar System \citep{wet92,caw01,pmc01,omb07}. The later migration of the giant planets, suggested to be responsible for the Late Heavy Bombardment, would have caused a second phase of depletion and possibly of enhanced collisional evolution of the asteroid belt \citep{gea05}. The violent depletion processes acting in the early Solar System left then place to slower, secular depletion mechanisms. Chaotic diffusion into the resonances with the giant planets has been suggested to have caused a depletion of a factor 2 of the population of large (D $> 10-30$ km) asteroids \citep{mam10}. Previous studies, however, always assumed the formation time of the giant planets as negligible and the giant planets (generally Jupiter and Saturn) were introduced instantaneously and all at the same time \citep{pmc01,omb07}.
In our previous work (\citet{tmc11}, hereafter Paper I), we suggested instead that Jupiter could have been the first giant planet to form and that its formation triggered a phase of primordial bombardment due to its rapid mass increase and its likely inward radial migration. This Jovian Early Bombardment (JEB in the following) preceded the depletion phases discussed before and its duration was estimated of the order of $0.5-1$ Ma (see Paper I), but it was suggested \citep{cea11} that the later formation of Saturn could act to extend the duration of the JEB. After the end of the JEB and the formation of Saturn, the Solar System and the asteroid belt are expected to resume the evolutionary path described by the generally accepted dynamical and collisional scenarios (see \citet{wet92,caw01,pmc01,omb07} for the dynamical scenario and \citet{bea05a,bea05b} for the unified dynamical and collisional scenario).  The effects of the JEB, which in Paper I we investigated using Vesta and Ceres as case studies in view of the arrival of the NASA Dawn mission at Vesta, depend on the extent and the time-scale of the Jovian migration but also on the SFD of the primordial planetesimals. In Paper I we suggested that a Solar Nebula whose population of planetesimals was dominated by large (i.e. 100-1000 km in diameter) planetesimals or where Jupiter’s radial migration exceeded a few tenths of AU would prove an extremely hostile environment for the survival of Vesta and Ceres. It must be noted, however, that the assumptions on the erosion of Vesta and Ceres due to the JEB did not account for the re-accretion of the excavated material and thus overestimated the implications of the bombardment for the survival of the two target bodies. In this paper we address the implications of the JEB for the global evolution of the asteroid belt using a more complete set of target bodies and a more realistic evaluation of the cumulative collisional erosion.
\section{Dynamical and Physical Model}

In our simulations we reproduced the evolution of a template of the forming Solar System across a temporal window of $2$ Ma, located during the Solar Nebula phase (see e.g. \citet{cea11} and references therein) and centred on the moment the planetary core of Jupiter reached its critical mass and the planet started to accrete its gaseous envelope. Our template of the Solar System is composed by the Sun, the forming Jupiter, a disk of planetesimals and a set of target bodies whose perturbations on the nearby planetesimals are considered negligible.\\
The numerical model we used in our simulations is based on the set of equations we described in Paper I: as we anticipated, in this paper we improved the model by introducing a more realistic estimate of the cumulative erosive effects of the JEB. In the following we will summarise the set of equations and the methods we used to evaluate the intensity of the JEB: for the full derivation of the equations we refer the interested readers to Paper I.

\subsection{Target bodies}

In place of Vesta and Ceres we used as our case study in Paper I, in this work we used two sets of synthetic bodies of different sizes as the targets of the JEB, to obtain a clearer picture of its effects on the planetesimals populating the asteroid belt.\\
All members of each set were characterised by the same orbit: the first set of bodies was located on circular, planar orbits at $2.30$ AU from the Sun (i.e. midway between the inner edge of the asteroid belt and the $3:1$ resonance with Jupiter, Region A in the following) while the second set was located on circular, planar orbits at $2.65$ AU from the Sun (i.e. midway between the $3:1$ and the $5:2$ resonances with Jupiter, Region B in the following).\\ Each set was composed by five different target bodies, characterised respectively by a diameter of $100$ km, $200$ km, $500$ km, $1000$ km and $2000$ km. An average density $\rho=3.0\,g\,cm^{-3}$ was assumed for all target bodies in order to evaluate their masses and escape velocities from the surface.

\subsection{Jupiter's formation and migration}\label{jupiter}

The evolution of Jupiter during our simulations is divided into two phases: the growth of the planetary core to its critical mass and the accretion of the nebular gas to form a massive envelope. The phase of gas accretion can also be divided into two sub-phases, a first one where the Jovian mass increases exponentially and a second one where the Jovian mass asymptotically approaches its final value (see e.g. \citet{cmt10}).\\

\subsubsection*{Accretion of Jupiter}

At the beginning of our simulations, Jupiter is a planetary embryo with mass $M_{0}=0.1\,M_\oplus$ that grows to the critical mass $M_{c}=15\,M_\oplus$ in $\tau_{c}=10^{6}$ years. Since the total accretion time of Jupiter's core is the sum of our $\tau_{c}$ to the time needed to form the initial Mars-sized planetary embryo, our choice of $\tau_{c}$ is consistent with the lower limits indicated by theoretical works for the formation of Jupiter's core (a few Ma, see e.g. \citet{nea07} and references therein). During this first phase, the mass growth of Jupiter is governed by the equation
\begin{equation}\label{coregrowth}
 M_{p}=M_{0}+\left( \frac{e}{e-1}\right)\left(M_{c}-M_{0}\right)\times\left( 1-e^{-t/\tau_{c}} \right)
\end{equation}
Upon the reaching of the critical mass value of $15\,M_\oplus$, the nebular gas surrounding Jupiter is assumed to become gravitationally unstable and to be rapidly accreted by the planet to form its massive envelope. During this phase, the mass growth of Jupiter is instead governed by the equation
\begin{equation}\label{gasgrowth}
 M_{p}=M_{c}+\left( M_{J} - M_{c}\right)\times\left( 1-e^{-(t-\tau_{c})/\tau_{g}}\right)
\end{equation}
where $M_{J}=1.8986\times10^{30}\,g=317.83\,M_{\oplus}$ is the final mass of the giant planet. The e-folding time $\tau_{g}=5\times10^3$ years is derived from the hydrodynamical simulations described in \citet{cm04}, \citet{lis09} and \citet{cmt10}.\\
As we anticipated, we followed the evolution of our template of the Solar System across the gas accretion phase for $\tau_{a}=10^{6}$ years, i.e. for $200\times\tau_{g}$.\\

\subsubsection*{Migration of Jupiter}

While Jupiter starts on a circular orbit in all our simulations, theoretical models indicate that forming giant planets should undergo Type I and II migrations and drift inward due to their interactions with the protoplanetary disk (see e.g. \citet{pea07} and references therein). The time-scale $\tau_{M}$ of migration is a non-linear function of the mass and the heliocentric distance of the forming planet: for a planet at $5.2$ AU, $\tau_{M}$ varies between $\sim10^{5}-10^{7}$ years for Type I migration and is of the order of $\sim10^{5}$ years for Type II migration \citep{dkh03}. As an approximation, we ignored the distinction between Type I and II migrations and started the migration of Jupiter as soon as its mass reaches $15\,M_\oplus$, which is equivalent to say that the value of $\tau_{M}$ becomes of the order of $\sim10^{5}$ years (ibid). As a consequence, Jupiter moves on a circular orbit while its planetary core is growing to the critical mass and starts to spiral inward once the phase of gas accretion begins. The equation governing the migration of Jupiter is functionally similar to Eq. \ref{gasgrowth} we used to describe the growth of the gaseous envelope:
\begin{equation}\label{radiallaw}
 r_{p}=r_{0}+\left( r_{J} - r_{0}\right)\times\left( 1-e^{-(t-\tau_{c})/\tau_{r}}\right)
\end{equation}
where $r_{0}$ is the orbital radius of Jupiter at the beginning of the simulation, $r_{J}$ is the final radius and $\tau_{r}=\tau_{g}=5\times10^{3}$ years. We assumed Jupiter's final semi-major axis equal to the present one, an assumption consistent with both the standard model of planetary formation and the scenario described by the first formulation of the Nice Model \citep{tsi05}. In our simulations we considered four different migration scenarios: $0$ AU (no displacement), $0.25$ AU, $0.5$ AU and $1$ AU. Since the parameter $\tau_{r}$ we used in Eq. \ref{radiallaw} is an e-folding time, displacements of $0.25$ AU, $0.5$ AU and $1$ AU with the assumed value of $\tau_{r}$ are equivalent to assuming values of $\tau_{M}$ at $5.2$ AU respectively of $\approx3.2\times10^{5}$ years, $\approx1.6\times10^{5}$ years and $\approx8\times10^{4}$ years, consistently with the results of theoretical studies \citep{pea07}.\\

\subsection{The disk of planetesimals}\label{disk}

The disk of planetesimals populating our template of the forming Solar System is simulated through a swarm of $6\times10^{4}$ massless particles.

\subsubsection{Dynamical characterisation of the planetesimals}

The disk of planetesimals extend between $2$ AU and $8$ AU ($2\,AU \leq a_{i} \leq 8\,AU$) and is assumed to be dynamically cold, i.e. the orbits of the planetesimals are characterised by low eccentricity ($0 \leq e_{i} \leq 3\times10^{-2}$) and inclination ($0\,rad \leq i_{i} \leq 3\times10^{-2}\,rad$) values.
The values of eccentricity and inclination associated to each massless particle were chosen randomly as
\begin{equation}
 e_{i}=e_{0} X,\,i_{i}=i_{0} (1-X)
\end{equation}
where $e_{0}=3\times10^{-2}$, $i_{0}=3\times10^{-2}$ rad and $X$ is a number extracted from a uniform distribution in the range $\left[0-1\right]$.\\
The dynamical evolution of the disk of massless particles was computed through a fourth order Runge-Kutta integrator with a self adjusting time-step. The time-step is chosen by evaluating at each given time the smallest time-scale $\tau_{min}$ between the orbital periods of Jupiter, of the target bodies and of the massless particles and the free-fall time of Jupiter-particle pairs considered as isolated systems. The time-step is then computed as
\begin{equation}
 t_{ts}=\tau_{min}/f_{ts}
\end{equation}
where $f_{ts}=100$ in our simulations.\\
While from a dynamical point of view the planetesimals populating the disk are treated as massless particles, we associated to each of them mass and density values in order to model the effects of their impacts on the target bodies. The mass values were obtained through the Monte Carlo methods we will describe in Sect. \ref{SFDs}, while for the density values we considered two compositional classes. Planetesimals formed in the inner Solar System ({\bf ISS} in the following) were considered rocky bodies with mean density $\rho_{ISS}=3.0\,g\,cm^{-3}$, while planetesimals formed in the outer Solar System ({\bf OSS} in the following) were considered volatile-rich bodies with mean density $\rho_{OSS}=1.0\,g\,cm^{-3}$. The change between the inner and the outer Solar System is assumed to coincide with the location of the Snow Line. When not explicitly stated otherwise, the Snow Line was assumed to be at $r_{SL}=4.0\,AU$ (see e.g. \cite{en08} and references therein).\\
Finally, we treated each massless particle as a swarm of real planetesimals by associating to each of them a normalization factor $\gamma$, described in the Sect. \ref{SFDs} and used to estimated a realistic number of impacts during the JEB basing on the method detailed in Sect. \ref{impacts}.
\subsubsection{Size-frequency distributions of the planetesimals}\label{SFDs}
We studied the effects of the JEB in disks characterised by different SFDs, which in turn link the JEB scenario to the different scenarios proposed for the formation and early collisional evolution of the primordial planetesimals. We used these SFDs to associated mass values to the test particles through the Monte Carlo approaches we will now describe.
\subsubsection*{Primordial planetesimals formed in a quiescent disk}
The first SFD we considered is that of a disk of planetesimals formed by gravitational instability of the dust in the mid-plane of a non turbulent protoplanetary nebula \citep{saf69,gaw73,wei80}. The protoplanetary nebula is assumed to have a mass of $M_{neb}=0.02$ M$_\odot$ distributed between $1-40$ AU with dust-to-gas ratio $\xi=0.01$ and density profile $\sigma=\sigma_{0}\left(\frac{r}{1\,AU}\right)^{-n_{s}}$, where $\sigma_{0}=2700$ g cm$^{-2}$ is the surface density at $1$ AU and $n_{s}=1.5$. For such a nebula it can be showed \citep{cfm81} that the average diameters of the planetesimals would follow the semi-empirical relationship
\begin{equation}\label{masslaw}
 \overline{m}_{p}=m_{0}\left( \frac{r}{1\,AU} \right)^{\beta}
\end{equation}
where $\overline{m}_{p}$ and $m_{0}$ are expressed in $g$, $r$ is expressed in $AU$ and $\beta=1.68$. The value $m_{0}$ is the average mass of a planetesimal at $1\,AU$, i.e. $2\times10^{17}$ g \citep{cfm81}. If we assume that the mass dispersion of the planetesimals about the average values of Eq. \ref{masslaw} is governed by a Maxwell-Boltzmann distribution, in Paper I we showed that we can associate a mass value to each test particle by means of a Monte Carlo method where the uniform random variable $Y$ varying in the range $[0,1]$ is 
\begin{equation}
 Y=\frac{2\gamma\left(3/2,y^{*}\right)}{\sqrt{\pi}}=P\left(3/2,y^{*}\right)
\end{equation}
where $P\left(3/2,y^{*}\right)$ is the lower incomplete Gamma ratio.
The inverse of the lower incomplete Gamma ratio can be computed numerically and, by substituting $y^{*}$ back with $m^{*}/\overline{m}_{p}(r)$ we obtain
\begin{equation}\label{massval}
 m(r)=\overline{m}_{p}inv\left(P\left(3/2,Y\right)\right)
\end{equation}
Since the use of massless particles assures the linearity of the processes investigated over the number of considered bodies, we can extrapolate the number of impacts expected in a disk of planetesimals by multiplying the number of impacts recorded in our simulations by a factor $\gamma$ where
\begin{equation}\label{ratio}
 \gamma=N_{tot}/n_{mp}
\end{equation}
where $n_{mp}=6\times10^{4}$ and $N_{tot}$ is given by
\begin{align}\label{ntot}
 N_{tot}=\int^{r_{max}}_{r_{min}} 2 \pi r n^{*}(r)dr= \nonumber \\
 =\pi^{3/2}\frac{\xi\sigma_{0}}{m_{0}}\left(1\,AU\right)^{2}\left(\frac{1}{2-n_{s}-\beta}\right)\times \nonumber \\
 \left(\left(\frac{r_{max}}{1\,AU}\right)^{2-n_{s}-\beta}-\left(\frac{r_{min}}{1\,AU}\right)^{2-n_{s}-\beta}\right)
\end{align}
where $r_{min}=2$ AU, $r_{max}=10$ AU and the symbol $1\,AU$ indicate the value of the astronomical unit expressed in cm, i.e. $1\,AU=1.49597870691\times10^{13}$ cm.\\
\subsubsection*{Primordial planetesimals formed in a turbulent disk}
The second SFD we considered is that of planetesimals formed by concentration of dust particles in low vorticity regions in a turbulent protoplanetary nebula \citep{chs08,chb10}.
Following \citet{cha10}, we assumed the protoplanetary nebula as characterized by a surface density $\sigma'_{0}=3500$ g cm$^{-2}$ at $1$ AU, a nebula density profile with exponent $n'_s=-1$, a Snow Line placed at $3.0$ AU (see Paper I) and a dust-to-gas ratio $\xi'=0.3$ beyond the Snow Line and $\xi'=0.15$ inside the Snow Line (see Fig. $14$, gray dot-dashed line, ibid). The results of \cite{cha10} supply the average diameter of planetesimals as a function of heliocentric distance (see Fig. $14$, gray dot-dashed line, ibid), from which we derived the following semi-empirical relationship analogous to Eq. \ref{masslaw}
\begin{equation}\label{chambers_mass}
 \overline{m'}_{p}=\frac{\pi}{6}\rho D_{0}^{3}\left( \frac{r}{1\,AU} \right)^{3\beta'}
\end{equation}
where $\beta'=0.4935$, $D_{0}=70$ km is the average diameter of the planetesimals at $1$ AU. By substituting the primed quantities to the original ones in Eqs. \ref{massval} and \ref{ntot}, we can thus obtain the mass and the normalisation factor for each massless particle through the same approach we described previously.\\
\subsubsection*{Planetesimals in the ``Asteroid were born big'' scenario}
The third and the fourth SFDs we considered are derived by the results of \citet{mea09}.
\cite{mea09} did not explore a specific model of planetesimal formation in quiescent or turbulent disks but instead tried to constrain the initial size-frequency distribution of planetesimals in the orbital region of the Main Asteroid Belt. Their results suggest that the best match with the present-day SFD of the Main Asteroid Belt is obtained for planetesimal sizes initially spanning $100-1000$ km (see Fig. $8$, ibid), a range consistent with their formation in a turbulent nebula. \cite{mea09} supplies two SFDs associated to this case: a first one describing the primordial SFD of the planetesimals, which spans $100-1000$ km (see Fig. $8a$, black dots, ibid), and a second, collisionally evolved one where accretion and break-up of the primordial planetesimals extended the size distribution between $5-5000$ km (see Fig. $8a$, black solid line, ibid). For each ISS impact event in our simulations, we then estimated the mass of the impacting planetesimal through a simple Monte Carlo extraction based on the cumulative probability distributions of the two SFDs supplied by \cite{mea09}. The normalization factor is estimated through Eq. \ref{ratio} using the total number of planetesimals in the asteroid belt supplied by the cumulative SFDs from Fig. $8a$ in \cite{mea09}. In using the SFDs from \cite{mea09} we did not considered the contribution of OSS impactors, since the authors only investigated the primordial SFD of planetesimals in the orbital region of the asteroid belt.

\subsection{Collisional evolution}\label{impacts}

To reproduce the collisional histories of the target bodies we opted for a statistical approach based on solving the ray--torus intersection problem between the orbital torus of a target body and the linearised path of a massless particle across a time step (see Paper I). 
Our method is similar to the analytical method developed by \cite{opi76}, but given that the Jovian perturbations may significantly change the orbits of the massless particles on time-scales analogous to their precession time-scales, we preferred the use of a numerical algorithm that did not require averaging over orbital angles other than the mean anomaly.\\
The orbital torus representing the spatial probability density of a target body is characterised by a mean radius $R_{T} = a_{A}$ and a section $\sigma_{T}= \frac{\pi}{4} \times \left( D_{A} f_{G} \right)^{2}$ where $a_{A}$ is the semi-major axis of the considered target body, $D_{A}$ is its physical diameter and $f_G$ is the gravitational focusing factor
\begin{equation}
 f_{G}=1+\left( \frac{v_{esc}}{v_{enc}}\right)^{2}
\end{equation}
with $v_{esc}$ being the escape velocity from the target body and $v_{enc}$ its relative velocity respect to the massless particle.\\
When a massless particle crosses a torus, the impact probability is the probability that both the particle and the target body will occupy the same spatial region at the same time. This is equivalent to writing
\begin{equation}
P_{coll}=\frac{min(\tau_{P},\tau_{A})}{T_{A}}
\end{equation}
where $T_{A}$ is the orbital period of the target body and $\tau_{A}$ and $\tau_{P}$ are respectively the time spent by the target body and the massless particle into the crossed region of the torus. Once the crossed region is identified by solving the ray--torus intersection problem (see Paper I), $\tau_{A}$ and $\tau_{P}$ can be derived in a straightforward way since the orbital velocities of the two bodies are known.\\
For the sizes and the densities we are considering, the escape velocity from the surface of the target bodies would be of the same order of magnitude of the ejection velocity of the excavated material. As a consequence, re-accumulation of the excavated material should be taken into account and the simple cratering approximation used in Paper I is not adequate to estimate the net erosion of the target bodies.\\
To evaluate the effective erosion cumulatively caused by the impacts, we need to estimate the amount of material excavated and ejected with velocities greater than the escape velocity from the surface of the target bodies. To estimate the net mass loss, we used the scaling law for rocky targets by \citet{hah07}:
\begin{equation}\label{h&hlaw}
\frac{m_{e}}{m_{i}}=0.0054\left(\frac{v_{i}\sin{\alpha}}{v_{e}}\right)^{1.65}\left(\frac{\rho_{i}}{\rho_{t}}\right)^{0.2}
\end{equation}
where $m_{e}$ is the escaped mass, $m_{i}$ is the mass of the impactor, $v_{i}$ is the impact velocity, $\sin{\alpha}$ is the impact angle relative to the normal to the target surface, $v_{e}$ is the escape velocity from the surface of the target body and $\rho_{i}$ and $\rho_{t}$ are the densities respectively of the impactor and the target body.\\
Since we are treating the collisional evolution of the target bodies in a statistical way, we are interested in the average erosion efficiency: therefore, we need to remove the dependency from the impact angle in Eq. \ref{h&hlaw}.
Following \citet{sve11}, we can average Eq. \ref{h&hlaw} over $\alpha$ to obtain \begin{equation}\label{svelaw}
\frac{m_{e}}{m_{i}}=0.03\left(\frac{v_{i}}{v_{e}}\right)^{1.65}\left(\frac{\rho_{i}}{\rho_{t}}\right)^{0.2}
\end{equation}
To get a more complete picture of the implications of the JEB for the collisional evolution of the asteroid belt, for every impact event we also computed the specific impact energy $Q_{D}$ expressed in units of the specific dispersion energy $Q^{*}_{D}$ of the target body. We evaluated the catastrophic disruption threshold $Q^{*}_{D}$ of the target bodies using Eq. $6$ from \citet{baa99} and the coefficients for basaltic targets computed by these authors (see Table $3$, ibid). We used the coefficients of the case $v_{i}=5\,km\,s^{-1}$ from \citet{baa99} for all impact events with a velocity greater or equal than $5\,km\,s^{-1}$ and those of the $v_{i}=3\,km\,s^{-1}$ for all the other impact events.
For those impacts where $0.01 \leq Q_{D}/Q^{*}_{D} < 1$, we substituted Eq. \ref{svelaw} with Eq. $8$ from \citet{baa99} expressed in terms of the eroded mass: 
\begin{equation}\label{benzlaw}
\frac{m_{e}}{m_{t}}=0.5+s\left(\frac{Q_{D}}{Q^{*}_{D}}-1.0\right)
\end{equation}
where $s=0.5$ for $v_{i}<5\,km\,s^{-1}$ and $s=0.35$ for $v_{i} \geq 5\,km\,s^{-1}$. The effects of impacts with $Q_{D}/Q^{*}_{D} \geq 1$ are not accounted for in our evaluation of the eroded mass.\\
We used the number of cumulative impacts with $Q_{D}/Q^{*}_{D} \geq 0.01$ (high--energy impacts) and $Q_{D}/Q^{*}_{D} \geq 1$ (critical--energy impacts) together with the cumulative eroded mass estimated with Eqs. \ref{svelaw} and \ref{benzlaw} as the main parameters for our analysis.
\section{Intensity of the JEB across the asteroid belt}
\begin{figure*}
\centering
\includegraphics[width=\textwidth]{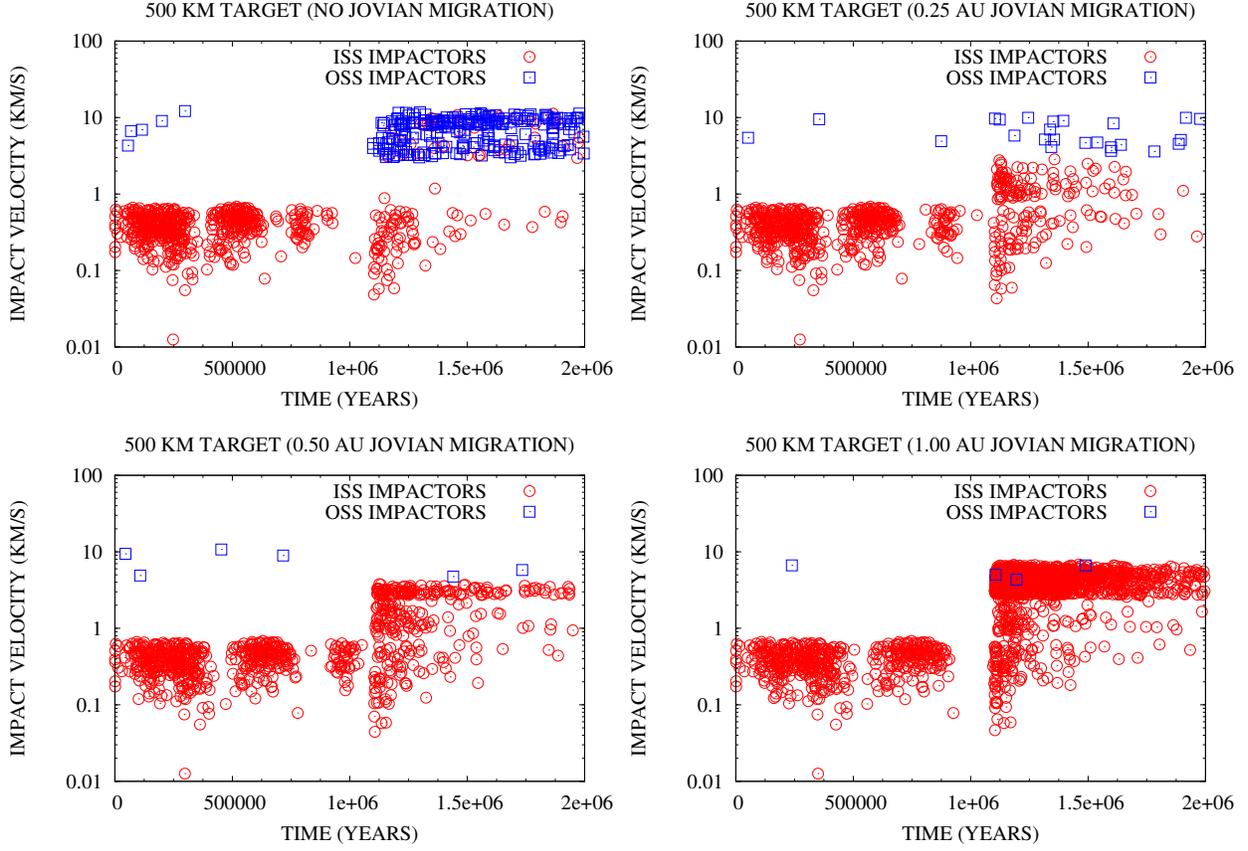}
\caption{Temporal evolution of the impact velocities of ISS (red symbols) and OSS (blue symbols) impactors on the target body with a diameter of 500 km in Region A in the four migration scenarios considered for Jupiter. The plots show the impacts events recorded in the simulations: the data are not normalised to the real disk population.}\label{fig1}
\end{figure*}

\begin{figure*}
\centering
\includegraphics[width=\textwidth]{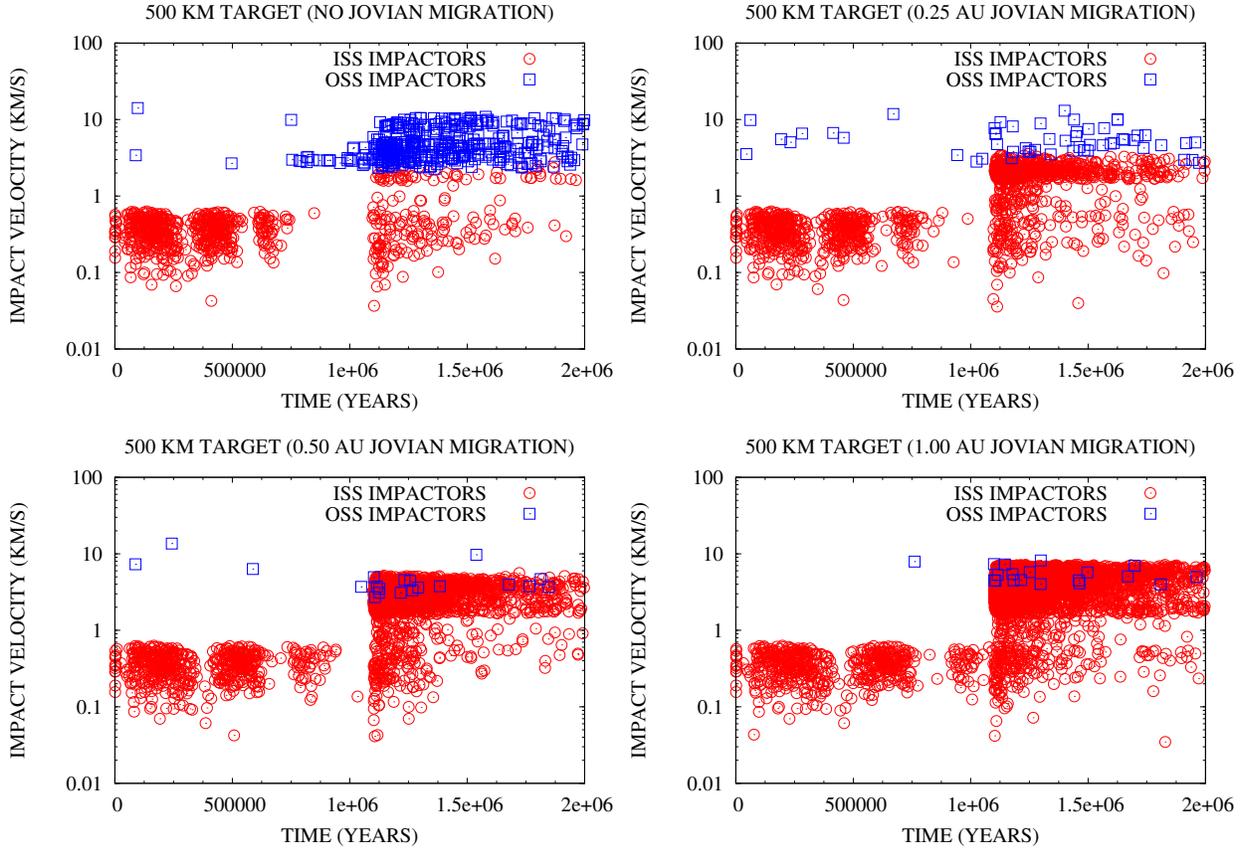}
\caption{Temporal evolution of the impact velocities of ISS (red symbols) and OSS (blue symbols) impactors on the target body with a diameter of 500 km in Region B in the four migration scenarios considered for Jupiter. The plots show the impacts events recorded in the simulations: the data are not normalised to the real disk population.}\label{fig2}
\end{figure*}

On the time-scales considered in our simulations, the flux of impactors associated to the JEB is dominated by the ISS planetesimals, in particular those affected by the $3:1$ and the $2:1$ resonances with Jupiter. The flux of OSS impactors is more erratic and, as we show in Figs. \ref{fig1} and \ref{fig2}, their contribution to the total flux is relevant mainly in the scenario where Jupiter does not migrate. As we show in Table \ref{fluxtable} for target bodies with radius of $100$ km, $200$ km and $500$ km, the location of a body with respect to the $3:1$ and the $2:1$ resonances with Jupiter and the extent of the Jovian displacement are the key factors in determining the intensity of the JEB for a given SFD of the planetesimals.\\ 
When Jupiter does not migrate the target bodies located in Region B receive a flux of impactors that is $130-150\%$ than that received by target bodies in Region A. When Jupiter migrates inward by $0.25$ or $0.5$ AU, target bodies located in Region B receive a flux that is a factor $200-300\%$ than that received by target bodies in Region A. In the case of planetesimals formed in a turbulent disk and characterized by the SFD from \citet{cha10}, the ratio increases to about $400-500\%$ (see Table \ref{fluxtable}). When Jupiter migrates by $1$ AU, however, the sweeping resonances are more efficient in enhancing the bombardment in Region A respect to the previous case. As a consequence, the flux of impactors in Region B ``drops'' again to $120-150\%$ respect to that in Region A.\\

\section{Planetesimal erosion during the JEB}
The implications of the phase of enhanced collisional activity associated to the JEB for the evolution of the asteroid belt strongly depend on the SFD of the planetesimals in the Solar Nebula at the time of the formation of Jupiter. Globally, the JEB causes the destruction of planetesimals up to a certain threshold size, whose exact value is a function of the extent of Jupiter's migration and of the abundance of large (i.e. $D \geq 100$ km) planetesimals in the Solar Nebula. Differently from what was generally assumed in previous studies of the collisional evolution of the asteroid belt \citep{bea05a,bea05b}, our results show that during the JEB the cumulative erosion due to non-critical impacts plays a role as important as, or possibly more important than, the one of critical impacts.\\
\subsubsection*{JEB in a quiescent disk}
If the first planetesimals formed in a quiescent Solar Nebula and the protoplanetary disk was still governed by its primordial SFD at the time Jupiter formed, the JEB would have important implications for the asteroids whose size varied between $100-200$ km (see Table \ref{quiescentdisk}). According to our results, critical impacts would cause the catastrophic disruption of planetesimals with size of about $100$ km in diameter in those scenarios where Jupiter migrated significantly (i.e. $1$ AU for target bodies in Region A and $0.5-1$ AU for target bodies in Region B). Moreover, due to the higher flux of OSS impactors, target bodies of about $100$ km in diameter would likely be shattered also in those scenarios where Jupiter did not migrate while forming (see Table \ref{quiescentdisk}). Planetesimals of $200$ km in diameter or bigger would not be undergo catastrophic collisions.\\
The cumulative erosion due to normal impacts ($Q_{D}/Q^{*}_{D} < 0.01$) and high-energy impacts ($0.01 \leq Q_{D}/Q^{*}_{D} < 1$), however, would cause the disruption of those bodies between $100-200$ km in diameter both in Region A and B in almost all migration scenarios (see Table \ref{quiescentdisk}). Even in the least erosive scenarios (Jupiter migrating by $0.25-0.5$ AU and target bodies located in Region A), planetesimals of $200$ km in diameter would lose between $25-75\%$ of their original mass due to collisional erosion. Collisional erosion would affect also planetesimals of about $500$ km in diameter in those scenarios where Jupiter migrates significantly (i.e. $1$ AU for Region A and $0.5-1$ AU for Region B), causing them to lose between $15-30\%$ of their original mass (see Table \ref{quiescentdisk}). For smaller displacements of the giant planet, planetesimals of about $500$ km in diameter would only lose a few per cent of their mass due to erosion.\\
Planetary bodies ranging $1000-2000$ km would be mostly unaffected by the JEB in such a protoplanetary disk.
\subsubsection*{JEB in a turbulent disk}
If planetesimal formation in the Solar Nebula was instead driven by turbulence and the protoplanetary disk was still characterized by its primordial SFD when Jupiter formed, critical impacts would dominate the collisional evolution of bodies as big as $200$ km in diameter (see Table \ref{turbulentdisk}), which would be destroyed both in Region A and B.\\
Planetary bodies of $500$ km in diameter or larger, however, would not undergo to critical impacts and their collisional evolution would be dominated by erosion. Planetesimals of $500$ km in diameter would be completely eroded both in Region A and B in almost all scenarios. Even in the only case where these planetesimals could survive the JEB (Jupiter migrating by $0.25$ AU and target bodies located in Region A), they would be stripped by about $30\%$ of their original mass. Also planetary bodies of $1000$ km in diameter located either in Region A or Region B would be completely eroded in those scenarios where Jupiter does not migrate or migrates by $1$ AU. In the other scenarios, the same bodies would be stripped of about $10-15\%$ of their original mass in Region A while in Region B the mass loss would rise to $30-50\%$ (see Table \ref{turbulentdisk}).\\
Finally, planetary embryos of about $2000$ km in diameter would lose only a few per cent of their original mass if the migration of Jupiter was limited (i.e. less than $1$ AU for Region A and less than $0.5$ AU for Region B) but their mass loss would rise to $5-10\%$ for higher values of the Jovian displacement (see Table \ref{turbulentdisk}).
%
\subsubsection*{JEB in the ``Asteroids were born big'' scenario}
In the most plausible scenario, i.e. a Solar Nebula populated by collisionally evolved  planetesimals, the survival of the primordial planetesimals would be in an intermediate position between the previous two cases, as is shown in Table \ref{evolveddisk}.\\
Planetesimals of about $100$ km in diameter would lose $50-100\%$ of their original mass both in Region A and Region B due to collisional erosion in all the considered migration scenarios. Also planetesimals of $200$ km in diameter would be completely eroded by the JEB for high values of the Jovian migration (i.e. $1$ AU for Region A and $0.5-1$ AU for Region B). In almost all the other scenarios, moreover, these planetesimals would be stripped by a significant fraction of their original mass due to collisional erosion ($10-30\%$ in Region A and about $50\%$ in Region B, see Table \ref{evolveddisk}).\\
Planetary bodies of $500$ km in diameter would be stripped by about $20-60\%$ of their original mass only in those scenarios where Jupiter migrates significantly (i.e. $1$ AU for Region A and $0.5-1$ AU for Region B). Planetesimals of $1000-2000$ km in diameter would instead survive the JEB without any significant mass loss.\\
Finally, the scenario described by the primordial SFD by \citet{mea09} is shown in Table \ref{primordialdisk} and is qualitatively analogous to the case of the JEB in a turbulent disk shown in Table \ref{turbulentdisk}. Planetesimals of $500$ km in diameter or smaller are either catastrophically disrupted or eroded by the JEB. Planetesimals of $1000$ km undergo a significant erosion ($10-50\%$ of their original mass) if the migration of Jupiter was limited (i.e. less than $1$ AU for Region A and less than $0.5$ AU for Region B) while are collisionally eroded in the other cases. Similarly, planetary bodies of $2000$ km in diameter suffer a limited erosion (a few per cent, see Table \ref{primordialdisk}) in most scenarios, but are stripped of $20-50\%$ of their original mass when Jupiter migrates significantly (i.e. $1$ AU for Region A and $0.5-1$ AU for Region B).\\
Before proceeding further, we want to stress once again that this estimate of the effects of the JEB in the ``Asteroids were born big'' scenario by \citet{mea09} takes into account only the contribution of ISS planetesimals. By comparison with the cases of planetesimals formed in  quiescent and turbulent disks and taking into account Figs. \ref{fig1} and \ref{fig2}, we can estimate that the inclusion of OSS impactors would likely imply the collisional erosion of bodies of $200$ km (collisionally evolved SFD) and $500$ km (primordial SFD) also in those scenarios were Jupiter does not migrate while forming.
%
\section{Caveats on the model and the results}
In applying the results we just described to assess the evolution of the early Solar System, one should be aware of the underlying assumptions and the present limitations of the model.
First, we assume that Jupiter was the first giant planet to form: this is equivalent to assume that the formation of Saturn took place a few $10^{5}$ years after that of Jupiter.
Second, we assume that the gravitational perturbations of Jupiter are the dominant effect on the dynamical evolution of the planetesimals across the JEB: we ignored the effects of gas drag and of the planetary embryos already formed in the Solar Nebula (see e.g. \citet{mea09,wei11}). The choice of not including neither gas drag nor planetary embryos is mostly due to the limitation of the numerical implementation of our model. As we stated previously, we use a fourth-order Runge-Kutta algorithm with a self-adapting time-step to reproduce the dynamical evolution of the disk of planetesimals. As we described in Paper I, we split the simulations into sub-simulations, each of them focusing on the dynamical evolution of a $1$ AU-wide ring of $10^{4}$ test particles. Each sub-simulation requires about $1$ month of computational time and we run a total of 48 simulations to assemble the dataset analysed in this work. The inclusion of gas drag, in particular, causes a significant increase in the impact events recorded in our simulations (please note we are not referring to the normalized flux of impactors) and the use of smaller time-steps due to the inward migration it forces on the planetesimals. With our choice of the numerical integrator, this translates in computational times in excess of $2$ months for some of the sub-simulations. We plan to explore the implications  of the inclusion of planetary embryos and gas drag in future works, using a more detailed version of the model and a more efficient N-Body algorithm.\\
Finally, the exclusion of the gravitational perturbations due to the presence of planetary embryos in the Solar Nebula implies that we do not take into account the depletion process of the asteroid belt \citep{wet92,caw01,pmc01,omb07} across the JEB. According to \cite{pmc01}, however, the presence of planetary embryos in the region of the asteroid belt does not cause a significant depletion of the planetesimals if Jupiter and Saturn are not present. This implies that the mass depletion process should be active only during the last $10^{6}$ years of our simulations. According to the results of \cite{omb07} for the case where Jupiter and Saturn are initially on circular orbits, the depletion of the asteroid belt on this temporal interval should amount to about $10\%$ of the original population of planetesimals. Since the bulk of the JEB takes place in  a few $10^{5}$ years (see Paper I), the depletion of the asteroid belt should not affect our results significantly.
\section{Discussion and Conclusions}
The results we presented here have some important implications, since they add new pieces to the mosaic of our understanding of the history of the early Solar System.\\ 
First, our results show that the formation of Jupiter triggers a short but intense primordial bombardment across the asteroid belt. The migration of Jupiter can act to enhance its intensity but is not necessary to start the Jovian Early Bombardment.\\
Second, our results clearly highlight the fact that, due to the more abundant population of the asteroid belt at the time of the Jovian Early Bombardment, cumulative erosion plays a more important role than that of critical impacts in determining the fate of the planetesimals. Such effects, not included in previous studies of the collisional history of the asteroid belt (e.g. \cite{bea05a,bea05b}), could help to explain the long equivalent time-scale ($10$ Ga instead of the real $4.5$ Ga, ibid) needed to achieve the degree of collisional evolution of the present-day asteroid belt.\\
Third, our results suggest that the generally accepted view that most asteroids of about $100$ km in diameter or larger are primordial may not be correct. We showed that the exact threshold size depends on the considered region of the asteroid belt, on the extent of Jupiter's migration, and on the size-frequency distribution of the planetesimals at the time of the Jovian Early Bombardment. We can generally state that, if the population of planetesimals in the Solar Nebula was dominated by objects smaller than $100$ km, the threshold size can be of the order of $200$ km. If, instead, the Solar Nebula was populated by planetesimals larger than $100$ km, the threshold size can rise up to about $500$ km. Planetesimals in the $100-200$ km size range are large enough to differentiate due to the energy released by the decay of short-lived radionuclides (see e.g. \citet{sco07} and references therein): the largest planetesimals disrupted by the Jovian Early Bombardment may therefore represent the parent bodies of the most ancient differentiated meteorites.\\
Fourth, our results support the claim we made in Paper I that the survival of Vesta to the Jovian Early Bombardment could be used to constrain the dynamical evolution of Jupiter. Our improved estimates of the collisional effects of the Jovian Early Bombardment indicate that, even in the least erosive scenarios, planetesimals of about $500$ km in diameter would be stripped of $20-40\%$ of their original mass if Jupiter migrated by about $1$ AU. Therefore, if Vesta formed near its present orbit its survival to the Jovian Early Bombardment suggests that Jupiter's migration was less than $1$ AU (see Paper I for a discussion).\\
Fifth, as we discussed in \citet{cea11} our results suggest that the formation of Saturn would trigger a second phase of bombardment, possibly less intense than the Jovian Early Bombardment due to the lower mass of Saturn and the depletion caused in the Solar Nebula by the formation of Jupiter. If the formation of Saturn took place only a few $10^{5}$ years after that of Jupiter, these two phases of bombardment would overlap into a single event, which in analogy to the Late Heavy Bombardment we called the Primordial Heavy Bombardment \citep{cea11}. The enhanced collisional activity of the inner Solar System caused by either the Jovian Early Bombardment or the Primordial Heavy Bombardment could prove extremely important to discriminate between the standard scenario for the evolution of the Solar Nebula we assumed in this work or alternative scenarios like the one recently proposed by \citet{wea11}.\\ 
Finally, since the physical processes considered here are general to planetary systems harbouring forming giant planets, our results indicate that the formation of an extrasolar giant planet in a circumstellar disk would trigger a phase of bombardment whose duration (i.e. about $0.5-1$ Ma) would be a significant fraction of the lifetime of the disk itself (i.e. about $10$ Ma, see e.g. \citet{mey08}). The erosive effects of such an extrasolar primordial heavy bombardment on already formed planetesimals would cause a sustained production of dust characterized by a large grain size (from tens of microns to cm, see e.g. \citet{jea07}). If the amount, the physical properties or the location of the dust produced by giant planet-induced bombardments differ significantly from those of the dust due to the collisional cascade caused by the presence of planetary embryos as observed in debris disks (see e.g. \citet{kri10} for a review), the enhancements in the dust population observed in circumstellar disks (see e.g. \citet{nea07} for a review) could in principle be used to identify those systems were the formation of extrasolar giant planets is taking place. The exploration of this possibility is the subject of ongoing research and will be discussed in a forthcoming paper.
\acknowledgements
Diego Turrini and Gianfranco Magni wish to dedicate this work to their friend and colleague Angioletta Coradini, who passed away after a long illness on September 5th, 2011. The authors also wish to thank the reviewer, John Chambers, for his comments on the manuscript and the results. This research has been supported by the Italian Space Agency (ASI) through the ASI-INAF contracts I/015/07/0 and I/010/10/0. The computational resources used in this research have been supplied by INAF-IAPS through the project ``HPP - High Performance Planetology''.

\begin{deluxetable}{ccccc}
\tabletypesize{\footnotesize}
\tablecolumns{5}
\tablecaption{Ratio of the flux of impactors caused by the JEB in Region B respect to that in Region A for the different SFDs and migration scenarios considered.\label{fluxtable}}
\tablehead{
\tableline
\colhead{Migration} & \colhead{Quiescent Nebula} & \colhead{Turbulent Nebula} & \colhead{Evolved SFD} & \colhead{Primordial SFD} \\
\colhead{Scenario} &  &  & \colhead{\citep{mea09}} & \colhead{\citep{mea09}}}
\startdata
 & & $100$ km target & & \\
\tableline
$0.00$ AU & $140.39$  &  $174.48$  &  $142.81$  &  $142.97$  \\
$0.25$ AU & $175.77$  &  $414.68$  &  $215.38$  &  $211.55$  \\
$0.50$ AU & $196.85$  &  $427.94$  &  $252.11$  &  $246.59$  \\
$1.00$ AU & $123.95$  &  $138.03$  &  $129.66$  &  $130.05$  \\
 & & $200$ km target & & \\
\tableline 
$0.00$ AU & $131.21$  &  $159.89$  &  $131.20$  &  $129.82$  \\
$0.25$ AU & $173.94$  &  $459.49$  &  $223.00$  &  $226.44$  \\
$0.50$ AU & $215.70$  &  $433.96$  &  $267.23$  &  $279.91$  \\
$1.00$ AU & $125.52$  &  $138.71$  &  $131.31$  &  $131.33$  \\
& & $500$ km target & & \\
\tableline
$0.00$ AU & $142.91$  &  $172.94$  &  $144.18$  &  $147.55$  \\
$0.25$ AU & $195.57$  &  $514.14$  &  $255.92$  &  $252.63$  \\
$0.50$ AU & $225.64$  &  $432.50$  &  $278.71$  &  $279.91$  \\
$1.00$ AU & $150.68$  &  $145.30$  &  $148.90$  &  $149.59$  \\
\enddata
\end{deluxetable}
\begin{deluxetable}{ccccccccccc}
\tabletypesize{\footnotesize}
\tablecolumns{9}
\tablecaption{Jovian Early Bombardment due to planetesimals formed in a quiescent disk following the SFD from \citet{cfm81}.\label{quiescentdisk}}
\tablehead{
 & & Region A & & & & Region B & & & \\
\tableline
\colhead{Migration} & \colhead{$N_{coll}$} & \colhead{High-energy} & \colhead{Critical} & \colhead{Eroded} & \colhead{$N_{coll}$} & \colhead{High-energy} & \colhead{Critical} & \colhead{Eroded}\\
\colhead{Scenario} & & \colhead{Impacts} & \colhead{Impacts} & \colhead{Mass} &  & \colhead{Impacts} & \colhead{Impacts} & \colhead{Mass}}
\startdata
 & & & & $100$ km & target & & & \\
\tableline
$0.00$ AU & $289.07$ & $12.17$  & $1.01$ & $1.79$ & $405.84$ & $41.36$  & $0.17$ & $2.64$  \\
$0.25$ AU & $396.42$ & $65.20$  & $0.01$ & $1.65$ & $696.77$ & $215.74$ & $0.04$ & $8.86$  \\
$0.50$ AU & $482.85$ & $101.24$ & $0.01$ & $5.02$ & $950.48$ & $382.84$ & $3.66$ & $29.18$ \\ 
$1.00$ AU & $940.55$ & $414.42$ & $14.63$& $55.14$& $1165.81$& $539.89$ & $18.92$& $67.48$ \\
 & & & & $200$ km & target & & & \\
\tableline 
$0.00$ AU & $653.45$ & $17.74$ & $0.00$ & $1.63$ & $857.40$  & $12.94$  & $0.00$ & $1.00$  \\
$0.25$ AU & $908.41$ & $1.52$  & $0.00$ & $0.25$ & $1580.11$ & $40.46$  & $0.00$ & $1.12$  \\
$0.50$ AU & $1083.75$& $45.09$ & $0.00$ & $0.74$ & $2337.68$ & $253.01$ & $0.00$ & $4.39$  \\
$1.00$ AU & $2300.67$& $515.34$& $0.00$ & $27.30$& $2887.85$ & $559.51$ & $0.00$ & $47.45$ \\
 & & & & $500$ km & target & & & \\
\tableline
$0.00$ AU  & $2261.79$ & $0.00$ & $0.00$ & $0.02$ & $3232.33$  & $0.00$ & $0.00$ & $0.02$ \\
$0.25$ AU  & $2860.79$ & $0.00$ & $0.00$ & $0.01$ & $5594.85$  & $0.00$ & $0.00$ & $0.06$ \\
$0.50$ AU  & $3439.61$ & $0.00$ & $0.00$ & $0.03$ & $7761.00$  & $0.00$ & $0.00$ & $0.14$ \\
$1.00$ AU  & $6659.37$ & $0.00$ & $0.00$ & $0.26$ & $10034.35$ & $0.00$ & $0.00$ & $0.34$ \\
\enddata
\tablecomments{The columns labelled $N_{coll}$, \emph{High-energy Impacts} and \emph{Critical Impacts} respectively report the total number of impacts, the number of impacts with $ 0.01 \leq Q_{D}/Q^{*}_{D} < 1$ and with $Q_{D}/Q^{*}_{D} \geq 1$. The column labelled \emph{Eroded Mass} reports the erosion due to all non-critical impacts in units of the target mass. All values are obtained by averaging over $10$ Monte Carlo extractions of the masses of the impactors.}
\end{deluxetable}
\begin{deluxetable}{ccccccccccc}
\tabletypesize{\footnotesize}
\tablecolumns{9}
\tablecaption{Jovian Early Bombardment due to planetesimals formed in a turbulent disk following the SFD from \citet{cha10}.\label{turbulentdisk}}
\tablehead{
 & & Region A & & & & Region B & & & \\
\tableline
\colhead{Migration} & \colhead{$N_{coll}$} & \colhead{High-energy} & \colhead{Critical} & \colhead{Eroded} & \colhead{$N_{coll}$} & \colhead{High-energy} & \colhead{Critical} & \colhead{Eroded}\\
\colhead{Scenario} & & \colhead{Impacts} & \colhead{Impacts} & \colhead{Mass} &  & \colhead{Impacts} & \colhead{Impacts} & \colhead{Mass}}
\startdata
 & & & & $100$ km & target & & & \\
\tableline
$0.00$ AU  & $4.77$  & $2.12$ & $2.64$  & $0.38$ & $8.31$  & $2.44$ & $5.86$  & $0.53$ \\
$0.25$ AU  & $5.41$  & $2.31$ & $3.09$  & $0.46$ & $22.45$ & $3.10$ & $19.32$ & $0.68$ \\
$0.50$ AU  & $9.69$  & $2.76$ & $6.92$  & $0.53$ & $41.48$ & $3.94$ & $37.51$ & $0.90$ \\
$1.00$ AU  & $43.31$ & $3.54$ & $39.75$ & $0.69$ & $59.78$ & $3.73$ & $56.05$ & $0.86$ \\
 & & & & $200$ km & target & & & \\
\tableline 
$0.00$ AU & $12.61$  & $7.50$  & $3.96$  & $0.43$ & $20.16$  & $12.70$ & $6.24$  & $1.34$  \\
$0.25$ AU & $12.40$  & $9.64$  & $1.19$  & $0.84$ & $56.97$  & $38.29$ & $17.66$ & $7.01$  \\
$0.50$ AU & $25.11$  & $14.28$ & $9.34$  & $1.94$ & $108.95$ & $53.87$ & $53.49$ & $10.54$ \\
$1.00$ AU & $113.35$ & $33.14$ & $78.49$ & $6.40$ & $157.23$ & $62.34$ & $93.55$ & $12.74$ \\
 & & & & $500$ km & target & & & \\
\tableline
$0.00$ AU & $48.93$ & $20.56$  & $0.12$ & $2.06 $ & $84.62$  & $40.88$  & $0.03$ & $2.11 $ \\
$0.25$ AU & $39.56$ & $8.12$   & $0.00$ & $0.31 $ & $203.42$ & $114.12$ & $0.01$ & $2.13 $ \\
$0.50$ AU & $80.55$ & $42.37$  & $0.01$ & $1.12 $ & $348.37$ & $237.87$ & $0.01$ & $6.55 $ \\
$1.00$ AU & $347.50$& $282.36$ & $0.00$ & $20.03$ & $504.93$ & $373.37$ & $0.01$ & $28.73$ \\
\enddata
\tablecomments{The columns labelled $N_{coll}$, \emph{High-energy Impacts} and \emph{Critical Impacts} respectively report the total number of impacts, the number of impacts with $ 0.01 \leq Q_{D}/Q^{*}_{D} < 1$ and with $Q_{D}/Q^{*}_{D} \geq 1$. The column labelled \emph{Eroded Mass} reports the erosion due to all non-critical impacts in units of the target mass. All values are obtained by averaging over $10$ Monte Carlo extractions of the masses of the impactors.}
\end{deluxetable}
\begin{deluxetable}{ccccccccccc}
\tabletypesize{\footnotesize}
\tablecolumns{9}
\tablecaption{Jovian Early Bombardment due to planetesimals following the collisionally evolved SFD from \citet{mea09}.\label{evolveddisk}}
\tablehead{
 & & Region A & & & & Region B & & & \\
\tableline
\colhead{Migration} & \colhead{$N_{coll}$} & \colhead{High-energy} & \colhead{Critical} & \colhead{Eroded} & \colhead{$N_{coll}$} & \colhead{High-energy} & \colhead{Critical} & \colhead{Eroded}\\
\colhead{Scenario} & & \colhead{Impacts} & \colhead{Impacts} & \colhead{Mass} &  & \colhead{Impacts} & \colhead{Impacts} & \colhead{Mass}}
\startdata
 & & & & $100$ km & target & & & \\
\tableline
$0.00$ AU & $1722.40$& $18.05$  & $0.06$& $2.72$ & $2459.84$ & $3.72$   & $0.13$& $0.48$   \\
$0.25$ AU & $2401.77$& $5.34$   & $0.12$& $0.62$ & $5172.90$ & $50.26$  & $0.44$& $3.54$   \\
$0.50$ AU & $3138.36$& $14.95$  & $0.26$& $1.87$ & $7912.08$ & $533.58$ & $0.68$& $12.55$  \\
$1.00$ AU & $7962.99$& $1376.98$& $0.75$& $93.21$& $10325.10$& $2128.11$& $0.93$& $221.16$ \\
 & & & & $200$ km & target & & & \\
\tableline 
$0.00$ AU  & $3855.44$  & $0.88$ & $0.00$ & $0.13$ & $5058.36$  & $0.52$ & $0.05$ & $0.06$ \\
$0.25$ AU  & $5420.12$  & $0.71$ & $0.04$ & $0.16$ & $12086.62$ & $1.05$ & $0.46$ & $0.48$ \\
$0.50$ AU  & $7404.21$  & $0.73$ & $0.24$ & $0.26$ & $19786.15$ & $2.33$ & $1.24$ & $1.30$ \\
$1.00$ AU  & $20071.18$ & $3.20$ & $1.22$ & $2.30$ & $26355.52$ & $3.13$ & $1.70$ & $3.16$ \\
 & & & & $500$ km & target & & & \\
\tableline
$0.00$ AU  & $13082.84$ & $0.08$ & $0.00$ & $0.01$ & $18862.24$ & $0.03$ & $0.00$ & $0.01$ \\
$0.25$ AU  & $16942.56$ & $0.24$ & $0.00$ & $0.01$ & $43358.98$ & $1.91$ & $0.00$ & $0.06$ \\
$0.50$ AU  & $23396.39$ & $1.48$ & $0.00$ & $0.04$ & $65209.09$ & $4.67$ & $0.00$ & $0.22$ \\
$1.00$ AU  & $59462.51$ & $3.63$ & $0.01$ & $0.42$ & $88538.54$ & $5.48$ & $0.01$ & $0.60$ \\
\enddata
\tablecomments{The columns labelled $N_{coll}$, \emph{High-energy Impacts} and \emph{Critical Impacts} respectively report the total number of impacts, the number of impacts with $ 0.01 \leq Q_{D}/Q^{*}_{D} < 1$ and with $Q_{D}/Q^{*}_{D} \geq 1$. The column labelled \emph{Eroded Mass} reports the erosion due to all non-critical impacts in units of the target mass. All values are obtained by averaging over $10$ Monte Carlo extractions of the masses of the impactors.}
\end{deluxetable}
\begin{deluxetable}{ccccccccccc}
\tabletypesize{\footnotesize}
\tablecolumns{9}
\tablecaption{Jovian Early Bombardment due to planetesimals following the primordial SFD from \citet{mea09}.\label{primordialdisk}}
\tablehead{
 & & Region A & & & & Region B & & & \\
\tableline
\colhead{Migration} & \colhead{$N_{coll}$} & \colhead{High-energy} & \colhead{Critical} & \colhead{Eroded} & \colhead{$N_{coll}$} & \colhead{High-energy} & \colhead{Critical} & \colhead{Eroded}\\
\colhead{Scenario} & & \colhead{Impacts} & \colhead{Impacts} & \colhead{Mass} &  & \colhead{Impacts} & \colhead{Impacts} & \colhead{Mass}}
\startdata
 & & & & $100$ km & target & & & \\
\tableline
$0.00$ AU & $11.61$  & $4.56$  & $7.04$  & $0.99$  & $16.60$  & $4.80$  & $11.80$  & $1.22$\\
$0.25$ AU & $16.44$  & $4.54$  & $11.91$ & $1.07$  & $34.79$  & $5.58$  & $29.21$  & $1.35$\\
$0.50$ AU & $21.32$  & $5.25$  & $16.07$ & $1.16$  & $52.58$  & $7.28$  & $45.29$  & $1.92$\\
$1.00$ AU & $53.17$  & $6.48$  & $46.68$ & $1.43$  & $69.14$  & $6.51$  & $62.63$  & $1.89$\\
 & & & & $200$ km & target & & & \\
\tableline 
$0.00$ AU & $26.16$  & $24.85$  & $1.31$ & $1.19$  & $33.96$  & $30.64$  & $3.32$  & $1.80$\\
$0.25$ AU & $35.84$  & $31.62$  & $4.22$ & $3.00$  & $81.15$  & $44.60$  & $36.56$ & $4.40$\\
$0.50$ AU & $47.82$  & $33.35$  & $14.48$& $3.10$  & $133.86$ & $54.84$  & $79.02$ & $4.56$\\
$1.00$ AU & $134.89$ & $51.42$  & $83.47$& $4.69$  & $177.16$ & $59.10$  & $118.06$& $5.21$\\
 & & & & $500$ km & target & & & \\
\tableline
$0.00$ AU & $85.24$  & $9.33$   & $0.11$ & $0.64$  & $125.77$ & $23.14$  & $0.07$ & $0.59 $\\
$0.25$ AU & $113.00$ & $34.50$  & $0.06$ & $0.58$  & $285.47$ & $156.10$ & $0.34$ & $4.12 $\\
$0.50$ AU & $154.54$ & $69.46$  & $0.18$ & $2.41$  & $432.58$ & $273.68$ & $1.12$ & $12.30$\\
$1.00$ AU & $397.37$ & $284.91$ & $2.12$ & $26.03$ & $594.43$ & $392.86$ & $2.63$ & $35.57$\\
\enddata
\tablecomments{The columns labelled $N_{coll}$, \emph{High-energy Impacts} and \emph{Critical Impacts} respectively report the total number of impacts, the number of impacts with $ 0.01 \leq Q_{D}/Q^{*}_{D} < 1$ and with $Q_{D}/Q^{*}_{D} \geq 1$. The column labelled \emph{Eroded Mass} reports the erosion due to all non-critical impacts in units of the target mass. All values are obtained by averaging over $10$ Monte Carlo extractions of the masses of the impactors.}
\end{deluxetable}
\end{document}